\documentclass[manuscript,nonacm]{acmart}

\AtBeginDocument{%
  }

\usepackage{makecell}
\usepackage{subcaption}

\begin{document}

\title{HapticMatch: An Exploration for Generative Material Haptic Simulation and Interaction}

\author{Mingxin Zhang}
 \affiliation{
   \institution{The University of Tokyo}
   \city{Chiba}
   \country{Japan}
 }
\email{m.zhang@hapis.k.u-tokyo.ac.jp}
\authornote{These authors contributed equally to this work.}

\author{Yu Yao}
 \affiliation{
   \institution{The University of Tokyo}
   \city{Chiba}
   \country{Japan}
 }
\email{yao@ms.k.u-tokyo.ac.jp}
\authornotemark[1]

\author{Yasutoshi Makino}
 \affiliation{
   \institution{The University of Tokyo}
   \city{Chiba}
   \country{Japan}
 }
\email{yasutoshi_makino@k.u-tokyo.ac.jp}

\author{Hiroyuki Shinoda}
 \affiliation{
   \institution{The University of Tokyo}
   \city{Chiba}
   \country{Japan}
 }
\email{hiroyuki_shinoda@k.u-tokyo.ac.jp}

\author{Masashi Sugiyama}
  \affiliation{
   \institution{RIKEN AIP}
   \country{Japan}
 }
 \affiliation{
   \institution{The University of Tokyo}
   \city{Chiba}
   \country{Japan}
 }

\email{sugi@k.u-tokyo.ac.jp}

\renewcommand{\shortauthors}{Zhang and Yao et al.}

\begin{abstract}
High-fidelity haptic feedback is essential for immersive virtual environments, yet authoring realistic tactile textures remains a significant bottleneck for designers. We introduce HapticMatch, a visual-to-tactile generation framework designed to democratize haptic content creation. We present a novel dataset containing precisely aligned pairs of micro-scale optical images, surface height maps, and friction-induced vibrations for 100 diverse materials. Leveraging this data, we explore and demonstrate that conditional generative models like diffusion and flow-matching can synthesize high-fidelity, renderable surface geometries directly from standard RGB photos. By enabling a "Scan-to-Touch" workflow, HapticMatch allows interaction designers to rapidly prototype multimodal surface sensations without specialized recording equipment, bridging the gap between visual and tactile immersion in VR/AR interfaces.
\end{abstract}

\begin{teaserfigure}
  \includegraphics[width=\textwidth]{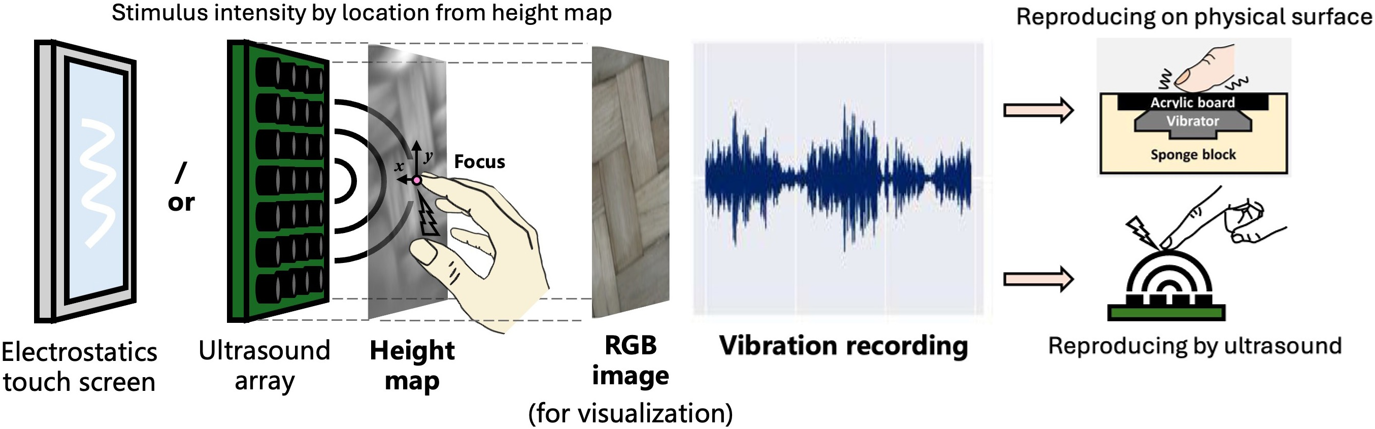}
  \caption{Electrostatic/Ultrasonic-based Height Maps Rendering}
  \label{fig:heisonic}
\end{teaserfigure}


\maketitle

\section{Introduction}
The digital simulation and spatial computing of human sensory information has seen remarkable success in vision and audition \cite{Szeliski_2011}, yet the simulation of haptic feedback \cite{10.1086/378619} remains the `missing link' in immersive experiences. In current digital environments, users can see the intricate weave of a fabric or the roughness of a stone wall, yet when they reach out to touch these surfaces on a touchscreen or interface, they feel only cold, flat glass. This sensory mismatch significantly breaks the illusion of presence, limiting the potential of VR for applications ranging from e-commerce to accessibility. A primary obstacle to bridging this gap is the haptic authoring bottleneck. Unlike visual textures, which can be easily captured with a standard camera, creating high-fidelity tactile assets for rendering hardware (such as electrostatic stimulators \cite{tanvas2022} or ultrasonic waves \cite{inoue2015active}) is prohibitively complex. Collecting this data, such as surface micro-geometry or interaction-induced vibrations, is traditionally time-consuming and resource-intensive. Consequently, the lack of easy-to-use tools prevents the widespread adoption of rich surface haptics in user interfaces. 

The recent success of diffusion-based generative models \cite{song2020denoising} in synthesizing high-fidelity images and audio \cite{karras2022elucidating, zhang2023survey} suggests a promising path. We envision a 'Scan-to-Touch' workflow: a paradigm where designers can upload a simple optical photo of a material, and the system automatically synthesizes the corresponding physical properties—specifically micro-height maps and vibration signals—needed for haptic rendering. Such a tool would allow rapid prototyping of tactile interfaces without requiring physical access to the materials or expensive recording setups.

\textbf{Our objective is to leverage fine-tuning and the generalization capabilities of existing pre-trained models to enable researchers in AR/VR haptic simulation \cite{bermejo2021survey} to rapidly batch-generate height maps for various common materials from optical images for testing and evaluation.}

To achieve this, a suitable dataset is critical. Existing haptic datasets, while valuable, often have limitations for multimodal generative tasks. For instance, the LMT108 dataset \cite{7737070} has been widely used \cite{10.1145/3275476.3275484} but lacks the aligned visual data necessary for visual-to-haptic generation. The more recent Touch and Go dataset \cite{10.5555/3600270.3600857} offers a broader range of materials but its unaligned data limits its use for synchronized feedback. Other large-scale datasets like OBJECTFOLDER 2.0 \cite{gao2022objectfolder} are tailored more for robotic manipulation than for fine-grained texture generation. Furthermore, a common workaround of using grayscale images as a proxy for height maps is unreliable, as lighting and color variations can lead to inaccurate geometric representations \cite{Fountouki20052025}.

To address these limitations, we introduce \textbf{HapticMatch}, a meticulously curated dataset designed to enable generative haptic interaction. Our contributions are threefold: 

First, we present the dataset comprising 100 common materials. For each material, we provide precisely aligned data pairs of (i) high-resolution optical surface images and (ii) corresponding micro-height maps from a GelSight-mini sensor \cite{yuan2017gelsight}, alongside (iii) vibrational audio signals from various interactions.

Second, We validate a pipeline using mainstream conditional generative models, including GANs \cite{goodfellow2020generative}, Flow-Matching \cite{liuflow} and Diffusion \cite{dhariwal2021diffusion}, demonstrating that it is possible to synthesize renderable height maps directly from visual inputs with high perceptual fidelity.

Third, we provide a proof-of-concept for AI-assisted haptic prototyping, establishing a benchmark that invites the HCI community to explore automated visual-to-tactile translation.

 The dataset is attached in supplementary materials and will be opensource.

\begin{figure*}[t]
  \centering
  \includegraphics[width=\textwidth]{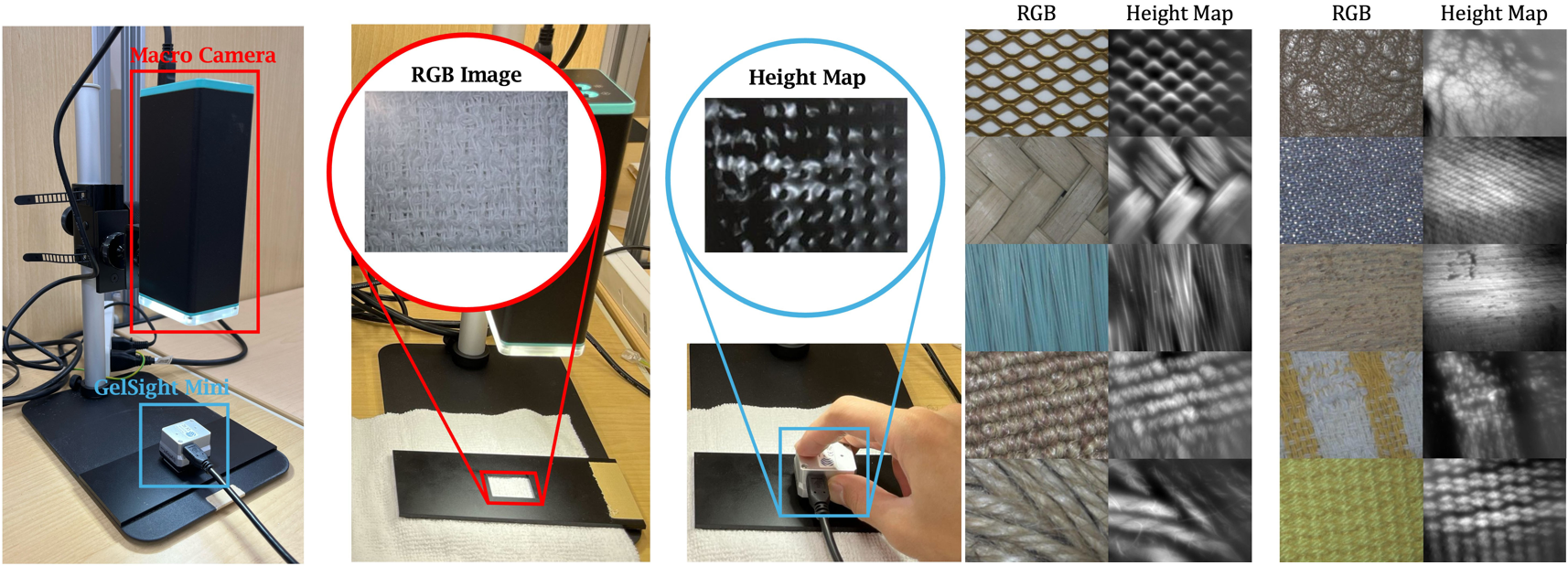}
  \caption{Visual-Haptic Data Acquisition (Left) and Data Samples (Right)}
  \label{fig:data}
\end{figure*}

\section{Datasets and Hardware}

\subsection{Aligned Visual-haptic images}
 The dataset covers five main categories—fabrics/leather, metals, plants, plastics, and other rigid materials such as rock and concrete, and each category consisting of 20 commonly encountered materials types. Each type in the dataset contains 20 pairs of $320\times240$ resolution images and spatially aligned height maps with totally 2000 data pairs. We prioritized relatively flat materials to helps avoid excessively uneven surfaces that may lead to failed contact with the surface. To facilitate broader usage, we also apply min-max normalization to the height maps, followed by scaling to the 0–255 range to produce grayscale images, which are also included in the dataset. Examples is shown in Fig. \ref{fig:data}. 
 
 Fig. \ref{fig:data} shows our setup for capturing paired RGB images and surface height maps. We constructed a data acquisition setup with a window to ensure that the RGB images captured by the camera and the height maps recorded by the GelSight sensor are precisely aligned in spatial position, scale, and orientation. 

 We did not constrain pressing force or use a robotic arm, allowing natural variability from human touch to introduce diverse interaction patterns. This promotes learning of more generalizable features. While applied force was not explicitly measured, it is implicitly captured through deformation patterns recorded by the GelSight sensor \cite{10.5555/3600270.3600857}.

\subsection{Vibration Audio Data}
For each material, we recorded 27 ten-second audio samples, varying the contact media (nail, finger, plastic stick), sliding speed (slow, medium, fast), and applied force (soft, medium, strong). For materials with anisotropic textures, we additionally recorded samples with the sliding motion parallel and perpendicular to the texture's primary direction. The data collection is manual for realistic effects by a fingertip-mounted directional microphone. The speed is controlled by electronic metronome, but the force label is only qualitative.




\subsection{Rendering Hardware}
Vibrotactile haptic rendering is largely limited to temporal acceleration cues, which are effective for transient events but lack spatial expressiveness \cite{10891204}. In contrast, electrostatic and ultrasonic haptic technologies enable spatiotemporal and localized tactile feedback \cite{mti6120108}, which is essential for texture rendering that relies on fine spatial variation and dynamic modulation of friction or pressure. Accordingly, we adopt generative models to support efficient batched data acquisition and scalable signal synthesis for texture rendering on advanced haptic hardware. Fig.~\ref{fig:heisonic} shows the demonstration.

\subsubsection{Electrostatic Haptic Display}
Electrostatic haptic displays, such as TanvasTouch \cite{tanvas2022}, exploit electro-adhesion to modulate friction at the contact surface. A time-varying electric field applied beneath the touch surface alters the effective normal force, thereby controlling lateral friction without mechanical actuation or surface deformation. As users slide their fingers across the surface, these friction modulations are perceived as fine surface textures. Owing to their high spatial resolution and rapid temporal response, electrostatic displays are well suited for rendering detailed texture maps and for integration with generative models that produce continuous spatiotemporal friction fields.

\subsubsection{Ultrasonic Haptic Display}
Ultrasonic haptic displays generate tactile sensations through acoustic radiation pressure produced by focused ultrasonic waves, enabling mid-air force feedback without physical contact \cite{inoue2015active}. By spatially controlling focal points, ultrasonic transducers can approximate surface geometry defined by height maps, while temporal modulation of acoustic parameters induces vibration cues associated with frictional and material properties \cite{10.1145/3476122.3484849}. Advanced modulation strategies further allow perceived pressure sensations beyond static acoustic limits \cite{9606531}, providing a flexible mechanism for reconstructing texture feedback from geometric representations.

\section{Models}
Given an optical image $I\in \mathbb{R}^{H\times W\times3}$ of a physical material surface, our goal is to generate a spatially aligned surface texture height map  
$\hat{T}\in \mathbb{R}^{H\times W}$ that is statistically and perceptually similar to the ground truth texture $T$.
We introduce general background of diffusion and flow-matching methods. The Conditional GAN is based on the setting in \cite{isola2017image}.
\subsection{Flow Matching}
Flow matching aims at constructing the probability path between the initial distribution $p_0$ and target distribution $p_1$, by learning a vector field $v_{\theta}: [0,1]\times\mathbb{R}^d\rightarrow\mathbb{R}^d$. Given the initial data $x_0 \sim p_0$, the generating process for target data can be described as evolving along an ordinary differential equation (ODE) $\frac{\mathrm{d}}{\mathrm{d}t}x_t=v_\theta(t,x_t)$ from $t=0$ to $t=1$ such that $x_1\sim p_1$. For usual generation tasks, $p_0$ is selected as a simple distributions such as Gaussian, but it is also possible to build the path between any two distributions. Following the work from \cite{liuflow}, in this paper, given data pairs $(x_0,x_1)\sim p_0\times p_1$, the target velocity for learning is selected as the straight line $u_t=x_1-x_0$, and the probability path is designed as the linear interpolation of two data points as $x_t=(1-t)x_0+tx_1$. The loss is defined as $\mathcal{L}_{\mathrm{FM}}(\theta)=\mathbb{E}_{t,(x_0,x_1)} \left[\| x_1-x_0-v_{\theta}(t,x_t) \|^{2} \right]$, where $\mathbb{E}$ is the expectation with respect to $t\sim\mathcal{U}[0,1]$ and $(x_0,x_1)\sim p_0\times p_1 $.
It is also possible to introduce extra variational autoencoder (VAE) structure $E(x)=z$ to implement in latent space \cite{dao2023flow}.
\subsection{Diffusion Model}
The Latent Diffusion Model (LDM) \cite{Rombach_2022_CVPR} first employs a pre-trained autoencoder with an encoder $\mathcal{E}$ and a decoder $\mathcal{D}$ to map images $x$ into a compressed latent space, where $z_0=\mathcal{E}(x)$. The forward process progressively adds Gaussian noise to the latent data $z_0$ over $T$ time steps according to a fixed variance schedule $\beta_1,\ldots,\beta_T$. A noisy latent $z_t$ at any timestep $t$ can be sampled in a closed form: $q(z_t|z_0)=\mathcal{N}\left(z_t;\sqrt{\bar{\alpha}_t}z_0,(1-\bar{\alpha}_t)\boldsymbol{I}\right)$, where $\mathcal{N}$ is the Gaussian, $\boldsymbol{I}$ is the identity matrix, $\alpha_t=1-\beta_t$, and $\bar{\alpha}_t=\Pi_{s=1}^{t}\alpha_s$. The reverse process will denoise $z_t$ back to $z_{t-1}$, which is parameterized by a neural network $\epsilon_{\theta}$ with the condition $c$. Instead of learning the full distribution $p_{\theta}\left(z_{t-1}|z_t\right)$, the model is trained to predict the noise component $\epsilon\sim\mathcal{N}(0,1)$ from the noisy input $z_t$ given a condition $c$. The training objective is formulated as  $\mathcal{L}_{\mathrm{LDM}}=\mathbb{E}\left[\| \epsilon - \epsilon_{\theta}\left( \sqrt{\bar{\alpha}_t}z_0+\sqrt{1-\bar{\alpha}_t}\epsilon,t,c\right) \|^2 \right].$

\section{Experiments}
To show the utility of our dataset, we implemented training on various img2img conditional generative models. Due to the limitation of the data scale, we trained simple models from scratch and large-scale model through fine-tuning for avoiding overfitting. To evaluate the performance quantitatively, we computed the average Learned Perceptual Image Patch Similarity (LPIPS) \cite{zhang2018unreasonable} score and compared the 2D Power Spectral Density (PSD) \cite{youngworth2005overview} between the ground truth and generated results. The GAN-based model and the flow-maching models based on UNet and DiT-B-2 architecture \cite{peebles2023scalable} are trained from scratch. The flow-maching models based on DiT-XL-2 \cite{peebles2023scalable} is trained by fine-tuning.

\begin{figure*}[t]
  \centering
  \includegraphics[width=\textwidth]{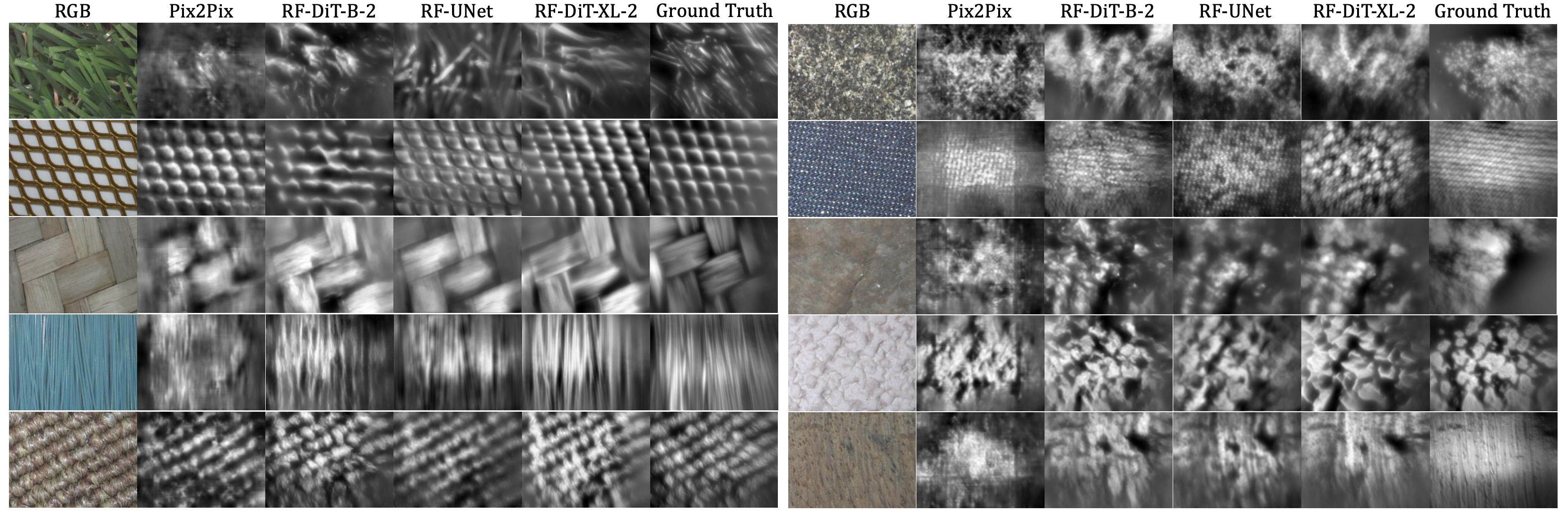}
  \caption{Visualized Examples of Different Generative Models Results}
  \label{fig:generation_results}
\end{figure*}

\subsection{Visual Results}
The visual example from different models is shown in Fig. \ref{fig:generation_results}. Pix2Pix exhibited significant limitations, primarily overfitting to simple visual cues like illumination and brightness rather than learning underlying 3D structures. Its outputs often resembled mere contrast enhancements and were marred by artifacts, likely due to its PatchGAN discriminator. A noticeable improvement was seen with RF-DiT-B-2, which began to capture genuine texture patterns. 

The RF-UNet also accurately reproduced most surface characteristics, though its outputs had lower contrast and softer edges, and it failed to render certain complex mesh-like patterns. Finally, the RF-DiT-XL-2 model demonstrated the best overall performance, consistently generating outputs with sharp, well-defined edges and detailed textures. Its only minor flaw was an occasional tendency to overfit to uneven color distributions in the input RGB image, resulting in less faithful reconstructions.

\subsection{Quantitative Results}
A common metric used to assess generative models, the Frechet Inception Distance (FID) \cite{heusel2017gans}, require large-scale datasets to produce reliable and meaningful evaluations \cite{heusel2017gans}. The limited scale of our dataset makes it challenging to apply FID as the metric in our case. Since the key feature of the generated height map is the local geometric structure of materials, we compute LPIPS and PSD between generated and ground truth height maps.

\subsubsection{LPIPS}
The overall results align broadly with subjective evaluations, though some discrepancies remain from Table \ref{tab:LPIPS}. Although the Pix2Pix model achieves superior LPIPS scores compared to RF-UNet, its generated outputs are of visibly lower quality than those from diffusion models. This suggests that LPIPS may prioritize structural or numerical similarity over authentic perceptual realism. Some samples of similar visual realism yielded disparate LPIPS scores, and textures with dense patterns often received worse LPIPS despite being perceptually close to their references. 
\subsubsection{PSD}
\begin{figure*}[t]
\centering
\begin{subfigure}[b]{0.245\textwidth}
\includegraphics[width=\textwidth]{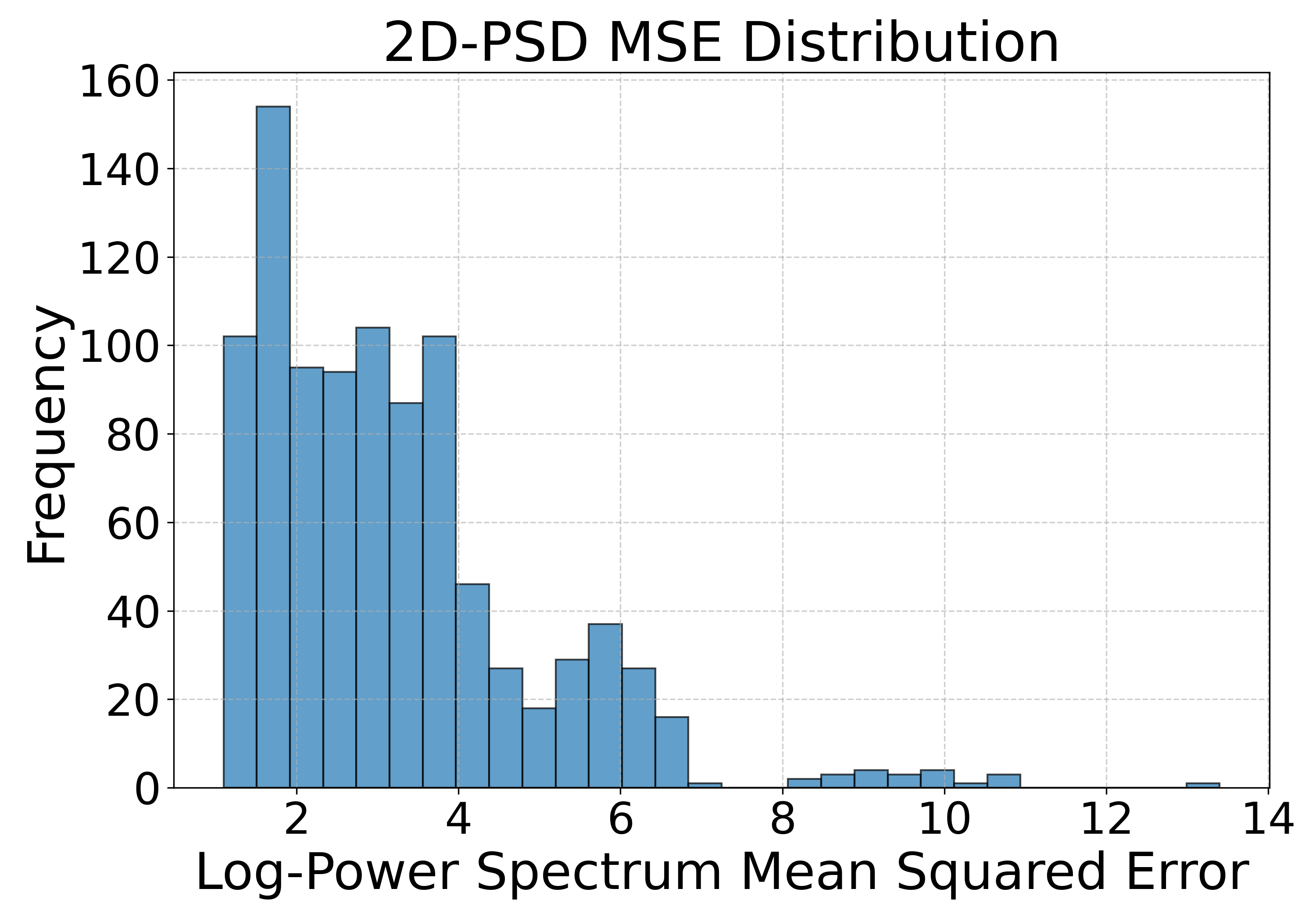}
\caption{RF-UNet}
\end{subfigure}
\begin{subfigure}[b]{0.245\textwidth}
\includegraphics[width=\textwidth]{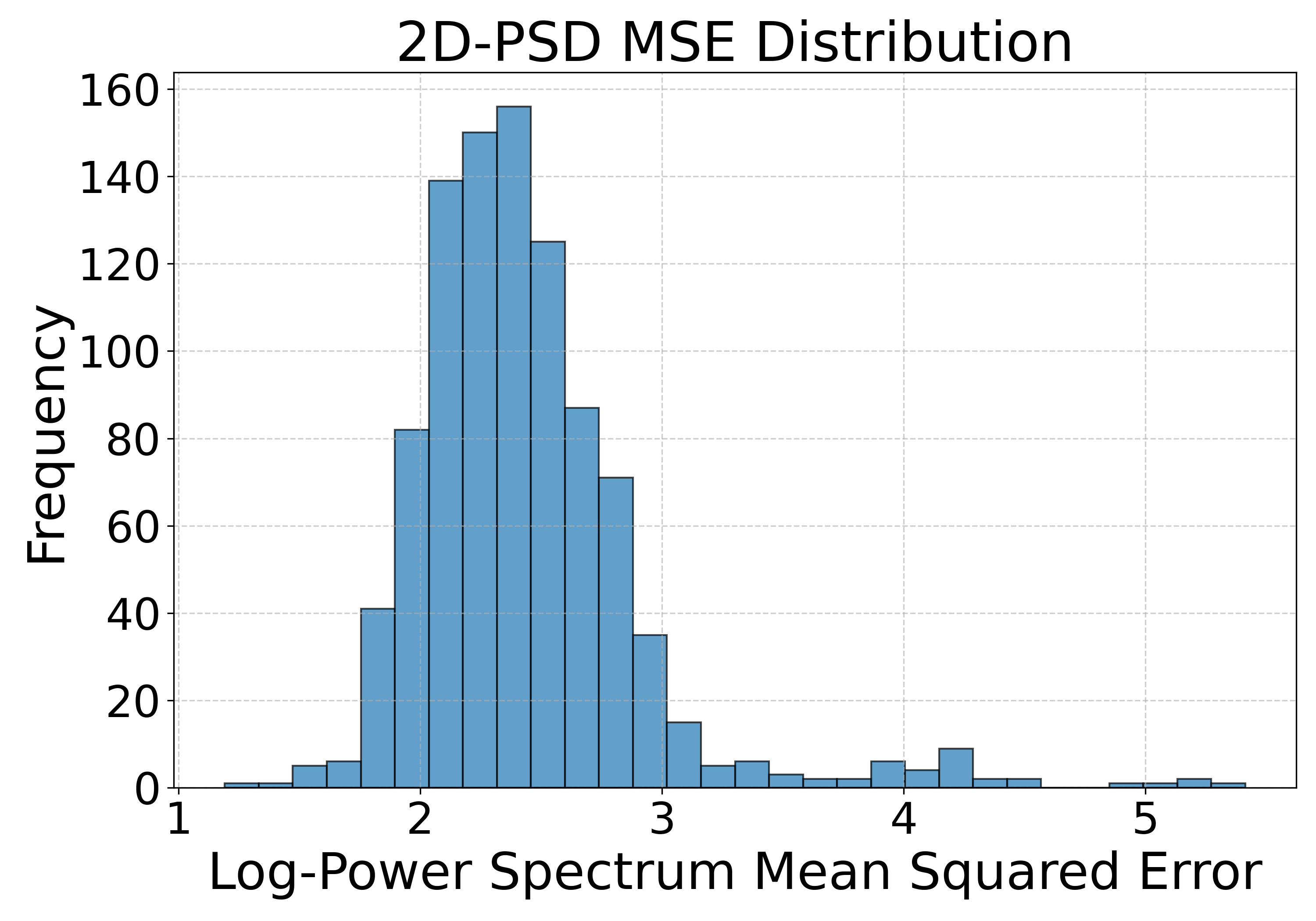}
\caption{Pix2Pix}
\end{subfigure}
\begin{subfigure}[b]{0.245\textwidth}
\includegraphics[width=\textwidth]{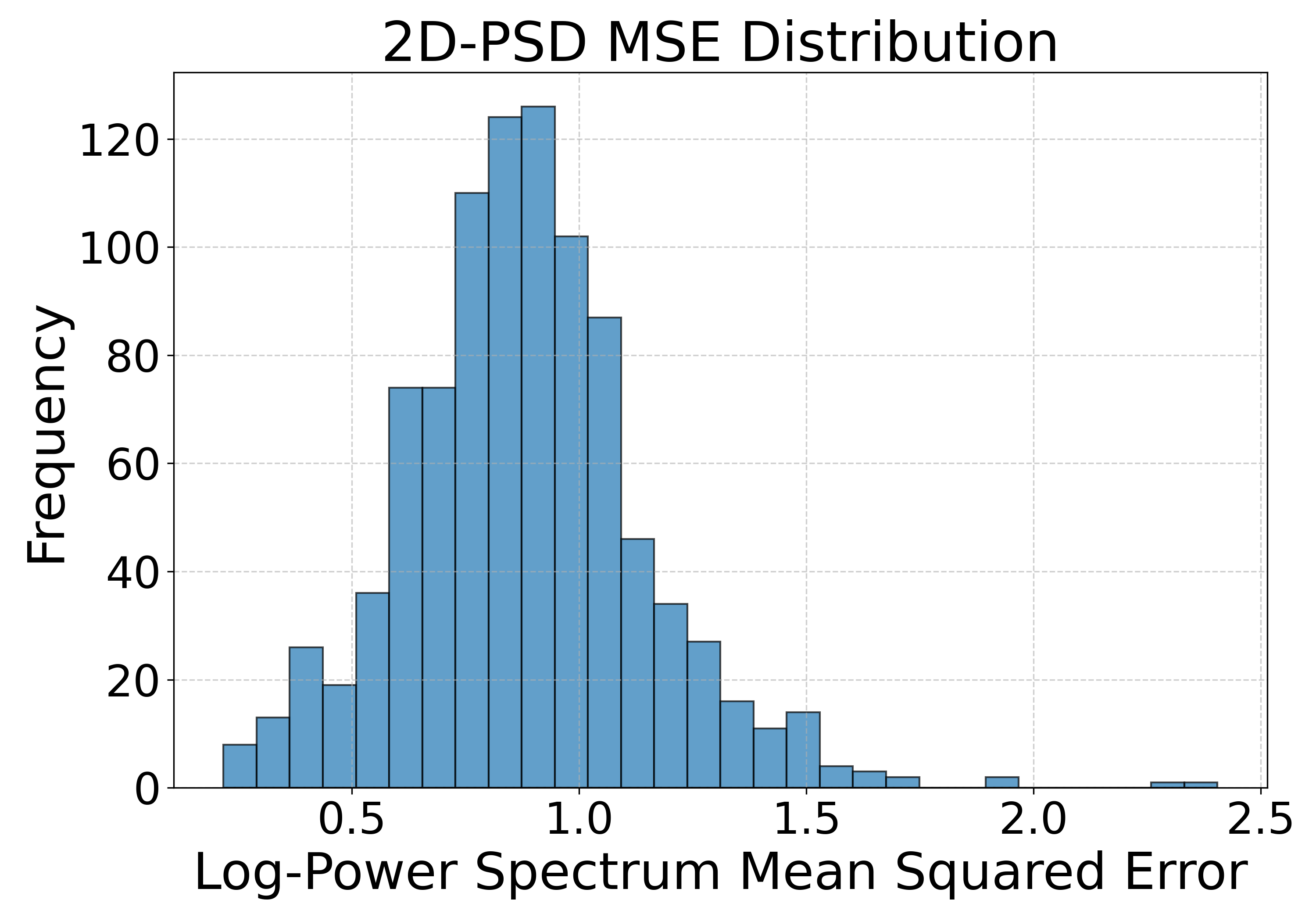}
\caption{RF-DiT-B-2}
\end{subfigure}
\begin{subfigure}[b]{0.245\textwidth}
\includegraphics[width=\textwidth]{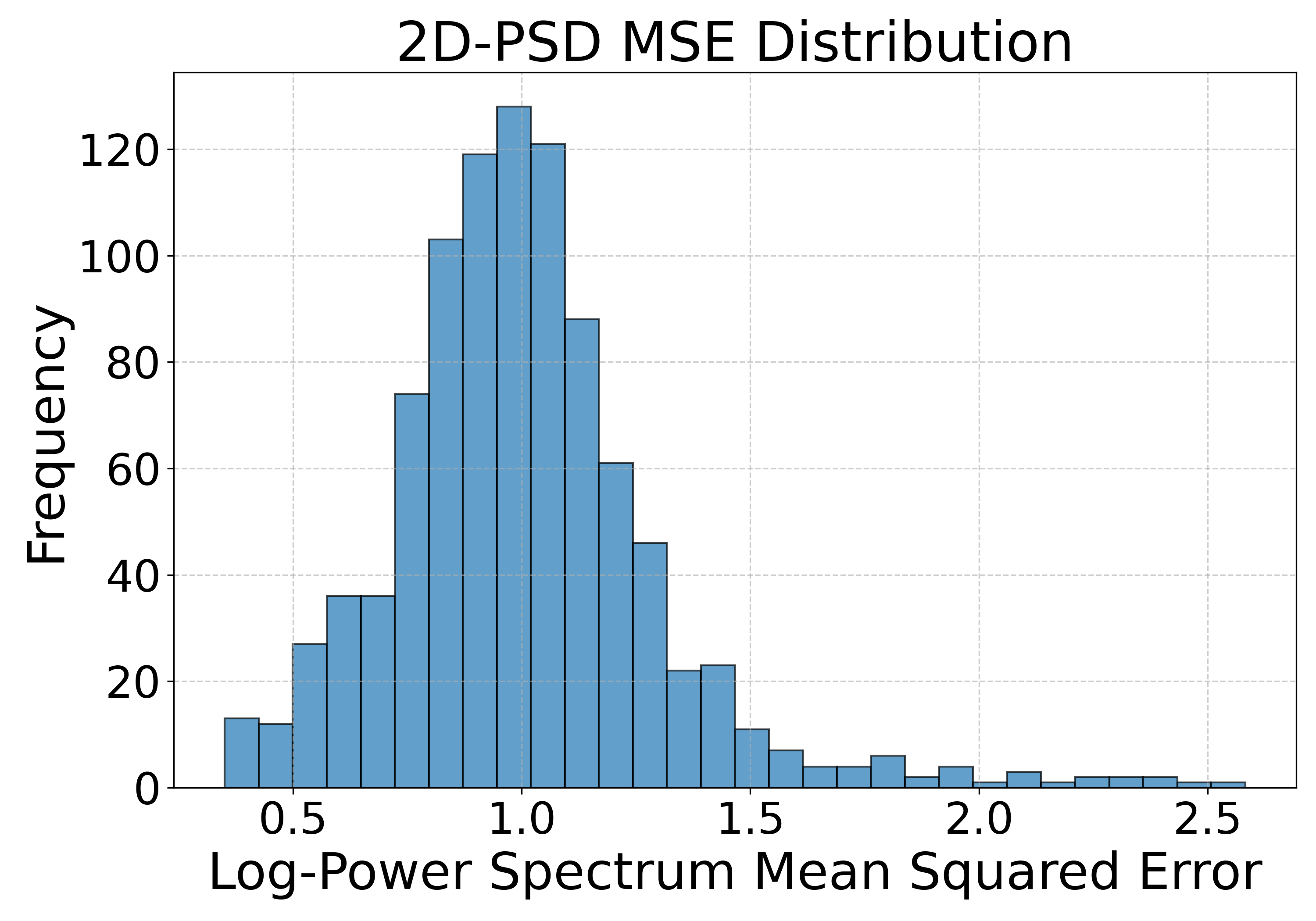}
\caption{RF-DiT-XL-2}
\end{subfigure}
\caption{The Log-PSD MSE Distribution of Evaluated Models}
\label{Log-distri}
\end{figure*}

The 2D Log Power Spectral Density (PSD) can capture key surface properties such as roughness, directionality, and periodicity by transforming the image from the spatial domain to the frequency domain. Compared to direct pixel-wise comparisons, this approach better reflects the perceptual similarity between textures.From Table \ref{tab:psd}, it can be observed that the performance of DiT-based Flow-matching methods is significantly better. The error distribution in Fig. \ref{Log-distri} also shows the general consistency in both two models.
\begin{table}[t]
    \centering
    \begin{minipage}[t]{0.49\linewidth}
        \centering
        \caption{The LPIPS of Evaluated ModelS.}
        \label{tab:LPIPS}
        \resizebox{\linewidth}{!}{
            \begin{tabular}{l l l}
            \toprule
            \textbf{Model} & \textbf{VGG} & \textbf{AlexNet} \\
            \midrule
            RF-UNet & \makecell[l]{0.58658\\(0.36621$\sim$0.72800)} & \makecell[l]{0.44261\\(0.20213$\sim$0.90576)} \\
            Pix2Pix & \makecell[l]{0.58376\\(0.34163$\sim$0.77945)} & \makecell[l]{0.39204\\(0.15962$\sim$0.74410)} \\
            RF-DiT-B-2 & \makecell[l]{0.43173\\(0.14862$\sim$0.67106)} & \makecell[l]{0.27764\\(0.04911$\sim$0.58818)} \\
            RF-DiT-XL-2 & \makecell[l]{0.49216\\(0.23856$\sim$0.68214)} & \makecell[l]{0.35562\\(0.14449$\sim$0.67667)}\\
            \bottomrule
            \end{tabular}
        }
    \end{minipage}
    \hfill 
    \begin{minipage}[t]{0.49\linewidth}
        \centering
        \caption{The Average Log-PSD MSE of Evaluated Models.}
        \label{tab:psd}
        \resizebox{\linewidth}{!}{
            \begin{tabular}{l l}
            \toprule
            \textbf{Model} & \textbf{2D Log-Power Spectral Density MSE}\\
            \midrule
            RF-UNet & \makecell[c]{3.2051 (1.0977$\sim$13.401)}\\
            Pix2Pix & \makecell[c]{2.4356 (1.1918$\sim$5.4122)}\\
            RF-DiT-B-2 & \makecell[c]{0.8797 (0.2178$\sim$2.4038)}\\
            RF-DiT-XL-2 & \makecell[c]{1.0031 (0.3552$\sim$2.5799)}\\
            \bottomrule
            \end{tabular}
        }
    \end{minipage}
\end{table}

\section{Dicussion}
\paragraph{Implications for Design and Interaction}
Our work bridges the gap between visual and tactile experiences in digital environments. Currently, creating high-fidelity haptic assets requires specialized hardware and tedious manual tuning. HapticMatch offers a generative workflow that democratizes this process: designers can simply input an optical photo to synthesize pixel-aligned height maps and vibration signals. This "Scan-to-Touch" capability significantly accelerates prototyping for VR/AR applications, allowing designers to populate immersive worlds with realistic surface properties—such as the roughness of rock or the weave of fabric—without requiring physical access to the materials.
\paragraph{Potential for Accessibility}
Beyond entertainment, our vision-to-tactile translation framework holds promise for accessibility. By converting standard images into renderable tactile signals (e.g., for electrostatic displays or ultrasonic arrays ), our approach could enable visually impaired users to "feel" digital photographs on touchscreens, enhancing information access through multimodal interaction.
\paragraph{Future Work}
Our immediate goal is to scale up HapticMatch by developing an automated data acquisition pipeline to reduce manual effort. Moving forward, we plan to training multi-modal models combining with audio data, and shift our evaluation focus from numerical metrics (like LPIPS/PSD) to human-centered validation. We aim to conduct psychophysical user studies to assess the perceptual realism of the generated textures and investigate how varying levels of generation fidelity affect user immersion in VR. Finally, we plan to integrate our generative models into game engines (e.g., Unity) to support real-time haptic rendering for interactive applications.


\bibliographystyle{ACM-Reference-Format}
\bibliography{sample-base}

\appendix

\end{document}